\documentclass[amsmath,amssymb,onecolumn,superscriptaddress]{revtex4}

\usepackage{graphicx}% 
\usepackage{dcolumn}% 
\def\be{\begin{equation}}
\def\ee{\end{equation}}
\def\beq{\begin{eqnarray}}
\def\eeq{\end{eqnarray}}

\usepackage{bm}
\usepackage{graphicx}
\usepackage{dcolumn}
\usepackage{amsmath}
\usepackage[latin1]{inputenc}
\usepackage{graphicx, psfrag}
\usepackage{amssymb}
\usepackage[colorlinks=true, citecolor=blue, urlcolor = blue, linkcolor= red, bookmarks=true]{hyperref}
\usepackage{float}
\usepackage{mathtools}
\usepackage{amsmath}
\usepackage{amsfonts}
\usepackage{dcolumn}
\usepackage{hyperref}
\usepackage{subfigure}
\usepackage{pgfplots}
\usepackage{epstopdf}
\usepackage{booktabs}
\usepackage{orcidlink}
\usepackage{xcolor}

\begin{document}
\title{Hybrid star within $f(\mathcal{G})$ gravity}

\author{Pramit Rej \footnote{Corresponding author} \orcidlink{0000-0001-5359-0655} }
\email[Email:]{pramitrej@gmail.com, pramitr@sccollegednk.ac.in, pramit.rej@associates.iucaa.in}
 \affiliation{Department of Mathematics, Sarat Centenary College, Dhaniakhali, Hooghly, West Bengal 712 302, India}

\begin{abstract}\noindent
The purpose of this work is to investigate some interesting features of a static anisotropic relativistic stellar object composed of two different types of fluid distributions typically termed as quark matter (QM) and ordinary baryonic matter (OBM) together with Krori-Barua type (KB) {\em ansatz} in the regime of modified $f(\mathcal{G})$ gravity, where $\mathcal{G}$ being the Gauss-Bonnet invariant term. In order to explain the correlation between pressure and matter density for the quark matter distribution within the compact object, we have taken into consideration the well-known MIT bag equation of state (EoS) whereas there is a simple linear correlation between pressure and matter density for ordinary baryonic matter. Furthermore, using graphical representations for varying parameters, the physical credibility of our obtained solutions has been intensively examined by regularity checking of the metric coefficients and matter variables, energy conditions, mass function, and causality conditions. For these analyses, we consider a particular compact stellar candidate 4U 1538-52. Finally, we found that the resulting outcome depicts the viability of the considered hybrid stellar model.

\end{abstract}

\maketitle

\section{Introduction}
The scientific community generally agrees that our universe is currently in an accelerated phase. General relativity (GR) in its standard form cannot explain the fact of accelerated expansion without the addition of new terms or components known as dark energy. Since the foundation of acceleration phenomena, many theories have been put forth to explain the origin of dark energy. These theories range from modifications to GR to the cosmological constant and scalar fields. This has produced a brand new, enticing research platform. There are several references in the literature on modified gravity theories with unified inflation-dark energy \cite {Nojiri:2006ri, Nojiri:2010wj, Nojiri:2017ncd, Odintsov:2018nch}. We also have comparative references to observational data sets \cite{Capozziello:2007ec, Capozziello:2009nq, Harko:2022unn, Farrah:2023opk}. The cosmological constant, whose origin can be explained by the vacuum energy density, is the most widely accepted concept, despite the fact that it differs from the value predicted by quantum field theories. Altering the usual gravity law is an additional option. Several methods for performing such a modification to GR have been proposed, including modifying the Einstein-Hilbert (EH) action, which obviously modifies the usual Einstein field equations (EFE). These modifications can be made by introducing some generic functions of the Ricci scalar or combinations of scalar and tensorial curvature invariants proposed by many relativistic astrophysicists. This method is now recognized as a well-established terminology, and its formulations may serve as a useful roadmap to investigate the cause of cosmic accelerated expansion \cite{Bloomfield:2012ff, Joyce:2016vqv, Langlois:2018dxi, Bamba:2022jyz}(further references therein).\par 

Starobinsky proposed a model that describes an inflation scenario in 1980, in which he inserts a ${\mathcal {R}}^2$ term into the EH action \cite{Starobinsky:1980te}. But Nojiri and Odintsov proposed the first consistent results of an accelerating universe from $f({\mathcal {R}})$ gravity by introducing a more complex function of the Ricci scalar ${\mathcal {R}}$ in action \cite{Nojiri:2003ft}. Such modifications effectively clarify the cosmic background via cosmological reconstruction, and they can also be used as replacements for dark matter and dark energy \cite{Cembranos:2010qd}. There are several other modifications to GR in the literature to explore the dark source terms on the dynamical evolution of astrophysical objects, such as $f({\mathcal {R}}, {\mathcal {T}})$ theory \cite{Harko:2011kv, Yousaf:2016qsd} (${\mathcal {T}}$ is the trace of stress-energy tensor), $f({\mathcal {R}}, {\mathcal {T}}, {\mathcal {R}}_{\mu \nu}, {\mathcal {T}}^{\mu \nu})$ theory \cite{Odintsov:2013iba} etc. Astashenok et al.\cite{Astashenok:2013vza} proposed a stable neutron star model in $f({\mathcal {R})}$ gravity. Shamir and Rashid \cite{FarasatShamir:2022gou} chose the isotropic matter distribution and Bardeen's model for compact star to find feasible solutions for the Einstein-Maxwell field equations within the framework of the modified $f({\mathcal {R})}$ gravity theory. Malik et al. \cite{Malik:2023ghy} utilized the isotropic distribution in the $f({\mathcal {R}, \mathcal {T})}$ theory of gravity to highlight the effect of electric charge on static spherically symmetric stellar structures. Rashid et al. \cite{Rashid:2023hlz} presented a set of exact spherically symmetric solutions for characterizing the interior of a relativistic star within the framework of the $f({\mathcal {R}, \mathcal {T})}$ modified theory of gravity. Astashenok et al.\cite{Astashenok:2023ewh} investigated compact objects composed of dense matter and dark energy in GR and modified gravity. Shamir and Meer \cite{Shamir:2023ouh} studied compact relativistic structures using recently proposed ${\mathcal {R}}+ \alpha {\mathcal {A}}$ gravity model, where ${\mathcal {R}}$ is the Ricci scalar and ${\mathcal {A}}$ is the anticurvature scalar. They investigated a new classification of compact stars embedded in class I solutions. Recently, Malik et al. \cite{Malik:2024zhb} investigated the charged anisotropic properties of compact stars within modified Ricci-inverse gravity by using the Karmarkar condition. Besides these works, numerous investigative works on modified gravity have been recorded in the literature, covering a variety of topics \cite{Rej:2021ngp, Bhar:2017xne, Rej:2021qpi, Bhar:2021iog, Bhar:2021oag, Rej:2023pge, Bhar:2021uqr, Das:2023bff, Bhar:2023hwu, Bhar:2023zwi, Karmakar:2023gci, Sharif:2023nsz, Banerjee:2023tmt, Kaur:2023gca}.\\
Among other modified gravity theories available in the literature, one is Gauss-Bonnet (GB) gravity, which has garnered the most attention among those \cite{Nojiri:2005jg, Cognola:2006sp, DeFelice:2008wz, Nojiri:2010wj} and is named $f(\mathcal{G})$ gravity, where $\mathcal{G}$ is the Gauss-Bonnet invariant term. It should be noted that the scalar $\mathcal{G}$ is a topological invariant in $(3+1)$ dimensional or lower spacetime. In order to prevent changes to the equations of motion, the Gauss-Bonnet term in the EH action must be incorporated. It is feasible to modify the field equations containing nonlinear terms in $\mathcal{G}$, and these are the $f(\mathcal{G})$ theories \cite{Rodrigues:2012qu, Houndjo:2013us, Astashenok:2015haa, Odintsov:2016hgc}, even if the linear term has no effect on the equations of motion. This theory has been widely applied to the study of the late-time accelerated expansion of the universe \cite{DeFelice:2010zz}. Recently, Rashid et al. \cite{Rashid:2023gtf} solved the Einstein-Maxwell field equations in the context of modified $f(\mathcal{G})$ gravity by employing the conformal killing vectors. Furthermore, it is observed that this $f(\mathcal{G})$ gravity is less restricted than $f({\mathcal {R}})$ gravity \cite{DeFelice:2008wz}.\\
Additionally, the study of finite-time future singularities and the acceleration of the universe during late-time epochs could greatly benefit from the application of $f(\mathcal{G})$ gravity \cite{Nojiri:2007bt, Bamba:2010wfw}. A number of fundamental cosmic issues, including inflation, late-time acceleration, and bouncing cosmology, have been addressed by Nojiri et al. \cite{Nojiri:2017ncd}. They also asserted that certain modified theories of gravity, such as $f({\mathcal {R}})$, $f(\mathcal{G})$, and $f(\mathfrak{T})$ theories (where $\mathfrak{T}$ is the torsion scalar), could be a useful mathematical tool for examining the well-defined picture of our universe. Further constraints on $f(\mathcal{G})$ models could be obtained by examining the energy conditions (EC) \cite{Kung:1995he, Kung:1995nh, PerezBergliaffa:2006ni}.\\
Due to the success of the $f(\mathcal{G})$ theory, we are now interested in deciphering some of the open mysteries of cosmology. The current work represents a contribution that advances the ideas raised in the earlier sections. As a result, the goal of this study is to first rebuild a stellar model comprising quark matter (QM) and ordinary baryonic matter (OBM) in modified $f(\mathcal{G})$ gravity theory. The presence of QM in fluid distribution plays a crucial role in the formation of ultra-dense strange quark objects. Also, QM's presence makes it more complicated to get exact solutions for hybrid stars and there are no earlier works on hybrid stars in this $f(\mathcal{G})$ theory as well.
To construct this type of model, we have taken into account the well-known Krori-Barua (KB) {\em ansatz} \cite{krori}. The KB space-time comprises a well-behaved metric function that is fully free from singularities; this is the main justification for using the KB metric in this current work to obtain a physically legitimate solution to the Einstein field equations. This is an alternative metric that may be applied in certain situations to characterize space-time's geometry. Researchers hope to learn more about this space-time and its possible uses in comprehending cosmology, gravity, and other related phenomena. The goal is to go beyond the traditional framework of gravitational theories in order to gain a deeper understanding of them and possibly discover new information about the nature of space-time and the cosmos. According to a literature review, numerous researchers have employed this {\em ansatz} to investigate various properties of compact stars in GR or modified gravity \cite{Rahaman:2011cw, Kalam:2012sh, MonowarHossein:2012ec, Bhar:2014mta, Abbas:2015yma, Momeni:2016oai, Deb:2017rhd, Nashed:2023uvk} (and further references therein). Next, we examine the dynamical stability of the reconstructed model to see if it can account for some phases of the evolution of the universe and generate exact solutions for compact stars that are comparable to observational data.\\
The layout of this article is as follows: It is divided into seven sections. Section \ref{grav} covers the extended form of Gauss-Bonnet gravity, and we present a feasible $f(\mathcal{G})$ gravity model. We provide the fundamental field equations for $f(\mathcal{G})$ gravity in Section \ref{interior}. Section \ref{sol} describes the analytic solution of the field equations for the viable $f(\mathcal{G})$ model. The physical analysis of the present model is covered in Section \ref{phy} along with graphical representations. The stability and viability of the present model have been analyzed in Section \ref{stable}. The last section provides the concluding remarks of the paper. The geometricized unit system ($G = c = 1$) has been used throughout this paper.

\section{Modified $f(\mathcal{G})$ gravity}\label{grav}

In this section, we will discuss the extensive version of Gauss-Bonnet gravity along with the corresponding equations of motion. For modified $f(\mathcal{G})$ gravity, the usual EH action corresponding to the matter Lagrangian $\mathcal{S}_m$ can be expressed as following \cite{Nojiri:2009fq}:
\begin{eqnarray}
 \label{action}
	\mathcal{S}=\int d^{4}x\sqrt{-g}\left[ \frac{\mathcal{R}}{2\kappa^2} + f(\mathcal{G}) \right]+
\mathcal{S}_m (g^{\mu\nu},\psi),
\end{eqnarray}
where $\mathcal{R}$ is the 4-dimensional Ricci scalar, $\mathcal{G}$ is the GB) term, $g$ is the determinant of the metric $g_{\mu \nu}$ and the gravitational coupling constant $\kappa^2=8\pi G$ with $G$ being the usual Newtonian constant. The expression for the GB invariant term $\mathcal{G}$ is given by,
\begin{equation} \label{gb}
\mathcal{G} = \mathcal{R}^2 - 4\mathcal{R}_{\zeta \eta} \mathcal{R}^{\zeta \eta} + R_{\zeta \eta \psi \delta} R^{\zeta \eta \psi \delta}
\end{equation}  
with $\mathcal{R}$ is the curvature scalar, $\mathcal{R}_{\zeta \eta}$ is the Ricci tensor and $R_{\zeta \eta \psi \delta}$ is the Riemannian tensor. We suppose that the EH action (\ref{action}) is defined for a feasible functional form of $f(\mathcal{G})$ that is consistent with observational data in an accelerating universe from various observational constraints like solar system testing, Cassini experiments, and furthermore. We further assume that $f(\mathcal{G})$ is a continuous function of the argument $\mathcal{G}$ and has all higher-order derivatives $f^{n\geq 2}(\mathcal{G})$. Incorporating the matter action $\mathcal{S}_m$, we can define the usual stress-energy tensor of matter fields by the standard definition as, 
\begin{equation} \label{st}
{T}_{\mu \nu}=-\frac{2}{\sqrt{-g}}\frac{\delta\left(\sqrt{-g}\mathcal{S}_m\right)}{\delta g^{\mu \nu}}
\end{equation}
Now, varying the above EH action (\ref{action}) for the metric tensor $g_{\mu \nu}$ (considering as a dynamical variable), we derive the full set of modified field equations given by the following form \cite{Nojiri:2010oco}:
\begin{eqnarray}
\kappa^2 T_{\mu \nu} &=& \mathcal{R}_{\mu \nu}-\frac{1}{2}\mathcal{R} g_{\mu \nu}+ 8\Bigg[R_{\mu \rho \nu \sigma} + \mathcal{R}_{\rho \nu} g_{\sigma \mu} -\mathcal{R}_{\rho \sigma} g_{\nu \mu}- \mathcal{R}_{\mu \nu} g_{\sigma \rho}+ \mathcal{R}_{\mu \sigma} g_{\nu \rho}+\frac{\mathcal{R}}{2}(g_{\mu \nu} g_{\sigma \rho}-g_{\mu \sigma} g_{\nu \rho}) \Bigg] \nabla^{\rho}\nabla^{\sigma}f_{\mathcal{G}} \nonumber\\&& 
+(\mathcal{G} f_{\mathcal{G}}-f)g_{\mu \nu} \label{tt1}
\end{eqnarray}
where $f_{\mathcal{G} \mathcal{G} \mathcal{G} \dots {\rm (n~times)}}=\frac{d^n f(\mathcal{G})}{d \mathcal{G} ^n}$. Here we follow the convention for curvature tensors by writing the metric $g_{\mu \nu}$'s signature as $(+ - - - )$. Moreover, the Riemannian tensor and the covariant derivative for a given vector field are obtained by $R^{\sigma}_{\mu \nu \rho}=\partial_{\nu}\Gamma^{\sigma}_{\mu \rho}-\partial_{\rho}\Gamma^{\sigma}_{\mu \nu}+ \Gamma^{\omega}_{\mu \rho}\Gamma^{\sigma}_{\omega \nu} -\Gamma^{\omega}_{\mu \nu} \Gamma^{\sigma}_{\omega \rho}$ and $\nabla_{\mu}V_{\nu}=\partial_{\mu}V_{\nu}-\Gamma^{\lambda}_{\mu}V_{\lambda}$, respectively. The matter sector also complies with an additional conservation law $\nabla^{\mu}T_{\mu \nu}=0$.  

\subsection{Power law $f(\mathcal{G})$ gravity model}

In this current study, we have used the well-known power law model of $f(\mathcal{G})$ gravity as proposed by Cognola et al. \cite{Cognola:2006eg} expressed as,
\begin{eqnarray}\label{fg}
f(\mathcal{G}) &=& \alpha_1 \mathcal{G}^{n_1}
\end{eqnarray}
where $\alpha_1$ and $n_1(>0)$ are certain arbitrary parameters. This power law model is very suitable for observational data, and it is also useful in predicting the unification of early-time inflation and late-time cosmic acceleration \cite{Nojiri:2007bt}. From a cosmological perspective, the physical plausibility has been examined in several Refs. \cite{Cognola:2007vq, DeFelice:2008wz, Bamba:2010wfw}. Here, we intentionally choose $n_1 = 2$ for the sake of simplicity and visualization purposes.

\section{Interior space-time and field equations}\label{interior}

To describe the interior space-time of a static spherically symmetric compact object, here we consider the interior line element in the standard form as follows,
\begin{equation} \label{line1}
ds^{2}=e^{\alpha (r)}dt^{2}-e^{\beta (r)}dr^{2}-r^{2}(d\theta^{2}+\sin^{2}\theta d\phi^{2}),
\end{equation}
where $\alpha(r)$ and $\beta(r)$ are corresponding gravitational potential functions of radial coordinate $r$ only.
We can express the corresponding energy-momentum tensor for an anisotropic two-fluid matter configuration as,
 \begin{equation} \label{t1}
  \begin{rcases}
    \begin{aligned}
      T_0^0 &=\rho^{\text{eff}}= (\rho+\rho_q), \\
      T_1^1 &=-p_r^{\text{eff}} =-(p_r+p_q), \\
      T_2^2 &=T_3^3=-p_t^{\text{eff}}=-(p_t+p_q),\\
      T_0^1 &=T_1^0=0.
    \end{aligned}
  \end{rcases} \text{Energy-Momentum Tensor}
\end{equation}
where $\rho^{\text{eff}}=(\rho+\rho_q)$, $p_r^{\text{eff}}=(p_r+p_q$), and $p_t^{\text{eff}}=(p_t+p_q$) are the effective energy density and pressure components respectively.\\
Equations (\ref{t1}) provided above describe the nature of the anisotropic source distribution in the interior of the formation of compact objects, which is formed by two types of matter: ordinary baryonic matter (OBM) and quark matter (QM). Here $\rho,~ p_r$ and $p_t$ denote the matter-energy density, radial and transverse pressure components, respectively, of the OBM, whereas $\rho_q$ and $p_q$ represent the corresponding matter-energy density and pressure associated with the QM. \\
When we are working with modified $f(\mathcal{G})$ theory, it is necessary to determine the curvature invariants in terms of metric coefficients ($\alpha$ and $\beta$), which are given by,
\begin{eqnarray}
\mathcal{R} &=&  e^{-\beta } \Bigg[\alpha''+(\alpha' -\beta')\Bigg(\frac{\alpha'}{2}+ \frac{2}{r} \Bigg)+ \frac{2}{r} \Bigg] - \frac{2}{r^2},   \\
R_{\zeta \eta \psi \delta}R^{\zeta \eta \psi \delta} &=&  \frac{4}{r^4}-\frac{ 8e^{-\beta }}{r^4}+\frac{ e^{-2\beta }}{4r^2} \Bigg[4r^2 \alpha''^2 -2r^2 \alpha'^3 \beta' + r^2 \alpha'^4 + \alpha'^2 \Big( r^2 (4\alpha'' + \beta'^2) +8 \Big) -4r^2 \alpha' \alpha'' \beta' \nonumber\\&& +8\beta'^2 + \frac{16}{r^2}\Bigg]
\end{eqnarray}
Hence, GB invariant term $\mathcal{G}$ reduces to,
\begin{equation} \label{gb1}
\mathcal{G} = \frac{ 2e^{-2\beta }}{r^2} \Bigg[ (2\alpha'' + \alpha'^2)(1-e^{\beta }) + (e^{\beta } - 3) \alpha' \beta' \Bigg]
\end{equation} 
Now, by solving the modified EFEs (\ref{tt1}) for $f(\mathcal{G})$ gravity by the help of the equations (\ref{gb}), (\ref{st}) and (\ref{line1}), we finally obtain the following set of independent equations as follows:
\begin{eqnarray}
\kappa^2(\rho+\rho_q) &=& \frac{1}{r^2 e^{\beta}} (\beta' r +e^{\beta}-1) -\frac{8}{r^2 e^{2\beta}}\left(f_{\mathcal{G} \mathcal{G} \mathcal{G}}\mathcal{G}'^2 +f_{\mathcal{G} \mathcal{G}}\mathcal{G}''\right) (1-e^{\beta}) \nonumber\\&& + (\mathcal{G} f_{\mathcal{G}}-f(\mathcal{G}))-\frac{4}{r^2 e^{2\beta}}\beta' \mathcal{G}' f_{\mathcal{G} \mathcal{G}} (e^{\beta} -3)
    , \label{fe1}\\
\kappa^2 (p_r+p_q) &=& \frac{1}{r^2 e^{\beta}} (\alpha' r -e^{\beta}+1) -\frac{4}{r^2 e^{2\beta}}\alpha' \mathcal{G}' f_{\mathcal{G} \mathcal{G}} (e^{\beta} -3)-(\mathcal{G} f_{\mathcal{G}}-f(\mathcal{G})), \label{fe2}\\
\kappa^2 (p_t+p_q) &=& e^{-\beta}\left( \frac{\alpha'^{2}}{4}+\frac{\alpha''}{2}-\frac{\alpha' \beta'}{4}+\frac{\alpha'-\beta'}{2r}\right)+\frac{4\alpha'}{r e^{2\beta}}\left(f_{\mathcal{G} \mathcal{G} \mathcal{G}}\mathcal{G}'^2 +f_{\mathcal{G} \mathcal{G}}\mathcal{G}''\right)+\frac{2\alpha'^2 \mathcal{G}' f_{\mathcal{G} \mathcal{G}}}{r e^{2\beta}} \nonumber\\&& +\frac{2 \mathcal{G}' f_{\mathcal{G} \mathcal{G}}}{r e^{2\beta}} (2\alpha''-3\alpha')-(\mathcal{G} f_{\mathcal{G}}-f(\mathcal{G})) \label{fe3}
\end{eqnarray}
where prime $(^{\prime})$ denotes the derivative with respect to the radial coordinate `r' and $\kappa^2=8\pi$.
The relationship between pressure and matter density due to the QM configuration is described by the following MIT Bag EoS model\cite{Witten:1984rs, Cheng:1998na}
\begin{eqnarray}
p_q=\frac{1}{3}(\rho_q-4B_g)\label{mit}
\end{eqnarray}
where $B_g$ denotes the Bag constant. It mainly signifies the difference in matter density between the perturbative and non-perturbative QCD vacuums \cite{Mak:2003kw}. Chodos et al. derived its unit as MeV/$\rm{fm}^3$ \cite{Chodos:1974je}.\\
In addition to this, for OBM, assume that the radial pressure $p_r$ is proportional to the matter density $\rho$, i.e.
\begin{eqnarray}
p_r= \omega \rho \label{eos}
\end{eqnarray}
where $\omega$ denotes the EoS parameter ranging between $(0,~1)$ and $\omega \neq \frac{1}{3}$.

\section{Solution of field equations}\label{sol}

For our current study, we have taken into account the well-known gravitational metric potentials proposed by Krori and Barua \cite{krori} (known as KB {\em ansatz}) as,
\begin{eqnarray}\label{kb}
  \begin{rcases}
    \begin{aligned}
\alpha(r) &=& Y r^2+Z,\\
\beta(r) &=& X r^2
\end{aligned}
  \end{rcases} \text{KB Metric}
\end{eqnarray}
where $X$, $Y$, and $Z$ are certain arbitrary constants to be calculated later numerically from a smooth matching of interior and exterior spacetimes. $X$ and $Y$ have dimension $\rm{km}^{-2}$, while $Z$ is a dimensionless quantity. The considered metric potentials produce a non-singular viable stellar model, which will be discussed in the next section.\\
For our chosen $f(\mathcal{G})$ gravity model (\ref{fg}) by using the metric expressions (\ref{kb}), we solve the field equations (\ref{fe1})-(\ref{fe3}). Thus, we obtain the matter density ($\rho$) and pressure components ($p_r , p_t$) of OBM as:
\begin{eqnarray}
\rho &=& \frac{e^{-4 r^2 X}}{4 (-1 + 3 \omega) \pi r^6}
   \bigg[2 e^{4 r^2 X} r^4 (-1 + 8 B_g \pi r^2) + 
    e^{3 r^2 X} \Big(2 r^4 - r^6 (X - 3 Y)\Big) - 
    128 \alpha_1 Y \bigg\{-3 + r^2 (-9 X + 10 Y) + \nonumber\\&&
       2 r^4 (-10 X^2 + 7 X Y + Y^2) + 
       r^6 (42 X^3 - 59 X^2 Y + 12 X Y^2 + Y^3)\bigg\} + 
    128 \alpha_1 Y e^{r^2 X}
       \bigg\{-6 + r^2 (-13 X + 14 Y) + \nonumber\\&& r^4 (-22 X^2 + 10 X Y + 4 Y^2) + 
       r^6 (35 X^3 - 36 X^2 Y + 7 X Y^2 + 2 Y^3)\bigg\} - 
    128 \alpha_1 Y e^{2 r^2 X}
       \bigg\{-3 + 
       r^2 (X - Y) \Big(-4 -\nonumber\\&& 2 r^2 (2 X + Y) + r^4 (X - Y) (3 X + Y)\Big)\bigg\}\bigg],   \label{f1}\\
p_r &=& \frac{\omega e^{-4 r^2 X}}{4 (-1 + 3 \omega) \pi r^6}
   \bigg[2 e^{4 r^2 X} r^4 (-1 + 8 B_g \pi r^2) + 
    e^{3 r^2 X} \Big(2 r^4 - r^6 (X - 3 Y)\Big) - 
    128 \alpha_1 Y \bigg\{-3 + r^2 (-9 X + 10 Y) + \nonumber\\&&
       2 r^4 (-10 X^2 + 7 X Y + Y^2) + 
       r^6 (42 X^3 - 59 X^2 Y + 12 X Y^2 + Y^3)\bigg\} + 
    128 \alpha_1 Y e^{r^2 X}
       \bigg\{-6 + r^2 (-13 X + 14 Y) + \nonumber\\&& r^4 (-22 X^2 + 10 X Y + 4 Y^2) + 
       r^6 (35 X^3 - 36 X^2 Y + 7 X Y^2 + 2 Y^3)\bigg\} - 
    128 \alpha_1 Y e^{2 r^2 X}
       \bigg\{-3 + 
       r^2 (X - Y) \Big(-4 -\nonumber\\&& 2 r^2 (2 X + Y) + r^4 (X - Y) (3 X + Y)\Big)\bigg\}\bigg] \label{f2}\\
p_t &=& \frac{e^{-4 r^2 X}}{8 (-1 + 3 \omega) \pi r^6}
   \bigg[e^{4 r^2 X} r^4 \Big(-1 + \omega (-1 + 32 B_g \pi r^2)\Big) + 
    128 \alpha_1 Y e^{r^2 X}
       (2 \omega (-6 + r^2 X (-1 + r^2 X) (13 + 35 r^2 X)) + \nonumber\\&&
       r^2 (12 + r (3 - r X (-1 + r^2 X) (14 + r (3 + 4 r X))) + 
          \omega (-8 + r (-9 + r X (-22 + 
                   3 r (-3 + r X (-14 + r (3 +\nonumber\\&& 4 r X))))))) Y + 
       r^4 (-2 + 14 \omega + r^2 (4 + \omega (2 - 9 r) + 3 r) X + 
          6 (1 - 3 \omega) r^4 X^2) Y^2 + 
       2 r^6 (2 \omega + (-1 + 3 \omega) r^2 X) Y^3) \nonumber\\&& + 
    128 \alpha_1 Y (2 \omega (3 + r^2 X (9 + 20 r^2 X - 42 r^4 X^2)) + 
       r^2 (-10 + 
          r (-3 + 2 r X (-10 + r (-3 + r X (-2 + 3 r (3 + 8 r X))))) \nonumber\\&& +
           \omega (10 + 
             r (9 + 2 r X (16 + 
                   r (9 + r X (65 - 9 r (3 + 8 r X))))))) Y + 
       2 r^4 (1 - r^3 X (3 + 14 r X) + 
          \omega (-5 + 3 r^2 X (-4 \nonumber\\&& + r (3 + 14 r X)))) Y^2 - 
       2 r^6 (\omega + 2 (-1 + 3 \omega) r^2 X) Y^3) - 
    e^{3 r^2 X}
      r^4 (-1 - r^2 X + r^4 Y (-X + Y) + 
       \omega (-1 + \nonumber\\&& r^2 (5 X - 6 Y) + 3 r^4 (X - Y) Y)) - 
    256 \alpha_1 Y e^{2 r^2 X}
       (r^2 Y (1 + r^2 X + r^4 X (-X + Y)) + 
       \omega (-3 + r^2 (-4 X + Y) \nonumber\\&& - r^4 (4 X^2 + X Y - 2 Y^2) + 
          r^6 (X - Y) (3 X^2 + X Y - Y^2)))\bigg] .\label{f3}
\end{eqnarray}
Also, the matter density and pressure due to the QM are obtained as,
\begin{eqnarray} 
\rho_q &=& \frac{e^{-4 r^2 X}}{8 (-1 + 3 \omega) \pi r^6}
 \bigg[-e^{4 r^2 X} r^4 (-3 - 3 \omega + 32 B_g \pi r^2) + 
    3 e^{3 r^2 X} r^4 (-1 + \omega (-1 + 2 r^2 X) - 2 r^2 Y) - 
    384 \alpha_1 Y e^{r^2 X}
       (2 \omega (-6 \nonumber\\&& + r^2 X (-1 + r^2 X) (13 + 35 r^2 X)) + 
       r^2 (9 + \omega + 2 (3 + \omega) r^2 X - 
          3 (5 + 9 \omega) r^4 X^2) Y + 
       2 r^4 (1 + \omega + (3 - 2 \omega) r^2 X) Y^2 \nonumber\\&& + (1 + 
          \omega) r^6 Y^3) + 
    192 \alpha_1 Y (4 \omega (-3 + r^2 X (-9 - 20 r^2 X + 42 r^4 X^2)) + 
       r^2 (13 + \omega + 
          2 (9 + \omega) r^2 X - (63 + 47 \omega) r^4 X^2) Y \nonumber\\&& + 
       2 r^4 (1 + \omega - 3 (-3 + \omega) r^2 X) Y^2  + (1 + 
          \omega) r^6 Y^3) + 
    192 \alpha_1 e^{2 r^2 X}
      Y (r^2 Y (5 + 2 r^2 (X + Y) - r^4 (X - Y) (3 X + Y)) + \nonumber\\&&
       \omega (-12 + r^2 (-16 X + Y) + 
          r^6 (X - Y) (4 X - Y) (3 X + Y) + 
          2 r^4 (-8 X^2 + X Y + Y^2)))\bigg] ,    \label{f4}\\
p_q  &=& \frac{e^{-4 r^2 X}}{8 (-1 + 3 \omega) \pi r^6}
  \bigg[e^{4 r^2 X} r^4 (1 + \omega - 32 B_g \omega \pi r^2) + 
    e^{3 r^2 X} r^4 (-1 + \omega (-1 + 2 r^2 X) - 2 r^2 Y) - 
    128 \alpha_1 Y e^{r^2 X}
       (2 \omega (-6 + \nonumber\\&& r^2 X (-1 + r^2 X) (13 + 35 r^2 X)) + 
       r^2 (9 + \omega + 2 (3 + \omega) r^2 X - 
          3 (5 + 9 \omega) r^4 X^2) Y + 
       2 r^4 (1 + \omega + (3 - 2 \omega) r^2 X) Y^2 \nonumber\\&& + (1 + 
          \omega) r^6 Y^3) + 
    64 \alpha_1 Y (4 \omega (-3 + r^2 X (-9 - 20 r^2 X + 42 r^4 X^2)) + 
       r^2 (13 + \omega + 
          2 (9 + \omega) r^2 X - (63 + 47 \omega) r^4 X^2) Y \nonumber\\&& + 
       2 r^4 (1 + \omega - 3 (-3 + \omega) r^2 X) Y^2 + (1 + 
          \omega) r^6 Y^3) + 
    64 \alpha_1 e^{2 r^2 X}
      Y (r^2 Y (5 + 2 r^2 (X + Y) - r^4 (X - Y) (3 X + Y)) \nonumber\\&& + 
       \omega (-12 + r^2 (-16 X + Y) + 
          r^6 (X - Y) (4 X - Y) (3 X + Y) + 
          2 r^4 (-8 X^2 + X Y + Y^2)))\bigg] .  \label{f5}
\end{eqnarray}
The expressions for effective matter-energy density, effective radial and transverse pressure for our present model are obtained as
\begin{eqnarray} 
\rho^{\text{eff}} &=& \rho + \rho_q   \nonumber\\ 
&=& \frac{e^{-4 r^2 X}}{8 \pi r^6}
  \bigg[e^{4 r^2 X} r^4 + e^{3 r^2 X} r^4 (-1 + 2 r^2 X) + 
    64 \alpha_1 Y (-12 + r^2 (-36 X + Y) + 
       r^6 (3 X - Y) (56 X^2 + 3 X Y - Y^2)  \nonumber\\&&  + 
       2 r^4 (-40 X^2 + X Y + Y^2)) + 
    64 \alpha_1 e^{2 r^2 X}
      Y (-12 + r^2 (-16 X + Y) + r^6 (X - Y) (4 X - Y) (3 X + Y) + 
       2 r^4 (-8 X^2 \nonumber\\&& + X Y + Y^2)) - 
    128 \alpha_1 e^{r^2 X}
      Y (-12 + r^2 (-26 X + Y) + 2 r^4 (-22 X^2 + X Y + Y^2) + 
       r^6 (70 X^3 - 27 X^2 Y \nonumber\\&& - 4 X Y^2 + Y^3))\bigg] , \label{ef1}\\ \nonumber\\
p_r^{\text{eff}} &=& p_r + p_q  \nonumber\\ 
&=& \frac{e^{-4 r^2 X}}{8 \pi r^4}
  \bigg[-e^{4 r^2 X} r^2 + e^{3 r^2 X} (r^2 + 2 r^4 Y) - 
    64 \alpha_1 e^{2 r^2 X}
      Y^2 (5 + 2 r^2 (X + Y) - r^4 (X - Y) (3 X + Y)) + 
    128 \alpha_1 Y^2  e^{r^2 X}
      (9  \nonumber\\&& + 2 r^2 (3 X + Y) + r^4 (-15 X^2 + 6 X Y + Y^2)) - 
    64 \alpha_1 Y^2 (13 + 2 r^2 (9 X + Y) + 
       r^4 (-63 X^2 + 18 X Y + Y^2))\bigg] ,\label{ef2}\\ \nonumber\\
p_t^{\text{eff}} &=& p_t + p_q  \nonumber\\ 
&=& \frac{e^{-4 r^2 X}}{8 \pi r^4} \bigg[-64 \alpha_1 e^{2 r^2 X} Y^2 \Big(1 + r^2 (-X + Y)\Big)^2 - 
    e^{3 r^2 X} r^4 \Big(X + r^2 X Y - Y (2 + r^2 Y)\Big) + 
    128 \alpha_1 Y^2 e^{r^2 X}
       \Big\{-3 + \nonumber\\&&
       r (-3 + r X (-8 + r (-3 + r X (-5 + r (3 + 4 r X)))) + 
          r (4 + r^2 X (2 - 3 r (1 + 2 r X))) Y + 
          r^3 (1 + 2 r^2 X) Y^2)\Big\}\nonumber\\&& - 
    64 \alpha_1 Y^2 (-7 + 
       r (-6 + r X (-22 + r (-12 + r X (-71 + 12 r (3 + 8 r X)))) + 
          2 r (3 + r^2 X (9 - 2 r (3 + 14 r X))) Y \nonumber\\&& + 
          r^3 (1 + 8 r^2 X) Y^2))\bigg].  \label{ef3}
\end{eqnarray}

\subsection{Determination of metric constants from junction Conditions}\label{match}

For further analysis of model parameters, the values of $X$, $Y$, and $Z$ must be fixed. Therefore, in this section, we will look at a hypersurface $\Sigma$ that serves as the boundary for interior and exterior geometries. Whether the boundary surface is constructed using the star's internal or external geometry, its intrinsic metric will be identical. It demonstrates that the metric tensor components will be continuous over the boundary surface for any coordinate system. In order to solve the system of field equations with the restriction that the radial pressure $p_r = 0$ at $r = \bar{R}$ ($\bar{R}$ is the stellar radius), matching conditions for the interior metric (\ref{line1}) are necessary. 
So, we smoothly match our interior metric (\ref{line1}) to the exterior Schwarzschild metric presented as
\begin{eqnarray}
ds_{\Sigma}^{2} &=& \left(1 - \frac{2\mathcal{M}}{r}\right)dt^2 - \left(1 - \frac{2\mathcal{M}}{r} \right)^{-1}dr^2
 - r^2(d\theta^2+\sin^2\theta d\phi^2), \label{eq22}
\end{eqnarray}
where $\mathcal{M}$ is the stellar mass.\\
Now at the boundary $r= \bar{R}$, the continuity of the metric coefficients $g_{tt}$, $g_{rr}$ and $\frac{\partial g_{tt}}{\partial r}$ between the interior and exterior regions yields the following set of relations:
\begin{eqnarray}
1 - 2\tilde{\mathcal{U}} = e^{\alpha (\bar{R})} &=& e^{Y\bar{R}^2 + Z},\label{eq23}\\
1 - 2\tilde{\mathcal{U}} = e^{-\beta (\bar{R})} &=& e^{-X\bar{R}^2},\label{eq24}\\
\frac{\tilde{\mathcal{U}}}{\bar{R}} &=& \bar{R} Y e^{Y\bar{R}^2 + Z}.\label{eq25}
\end{eqnarray}
where, $\tilde{\mathcal{U}}=\frac{\mathcal{M}}{\bar{R}}$ is a dimensionless quantity.\\
\begin{figure}[H]
    \centering
        \includegraphics[scale=.55]{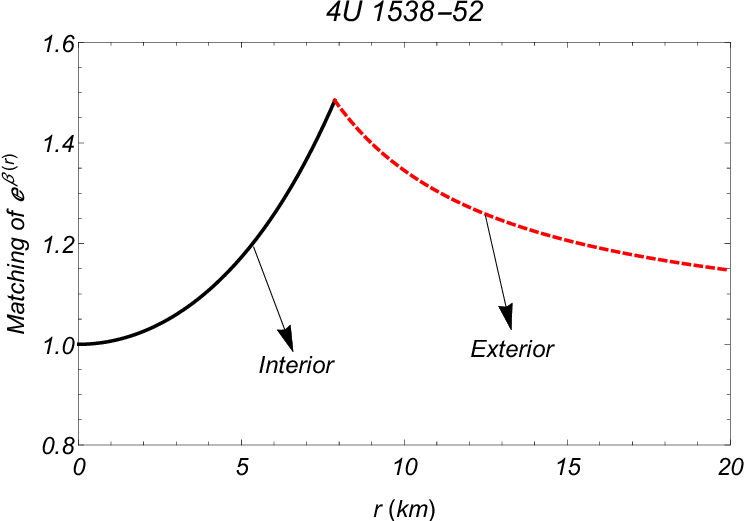}
         \includegraphics[scale=.55]{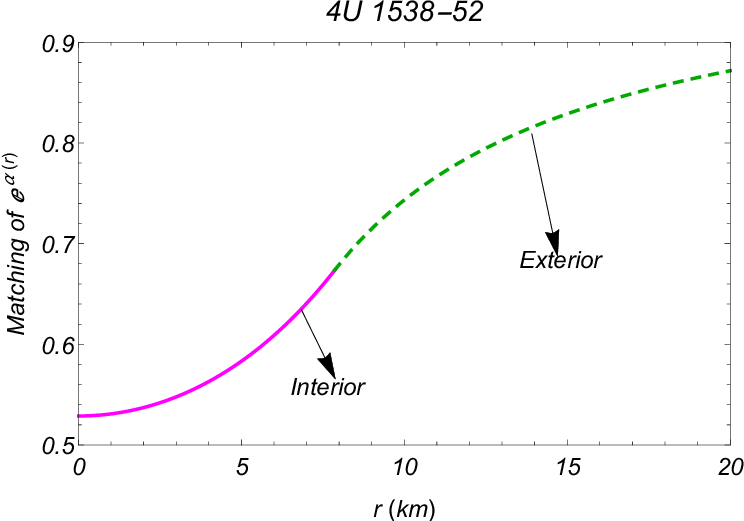}
        \caption{The matching condition of the metric potentials $e^{\beta(r)}$ and $e^{\alpha(r)}$ are shown against radius `r'.}\label{match}
\end{figure}
Now solving the expressions~(\ref{eq23})-(\ref{eq25}), we obtain the following values of the constants $X$, $Y$, $Z$ as
\begin{eqnarray}
X &=& -\frac{\rm{ln} (1 - 2\tilde{\mathcal{U}}) }{\bar{R}^2},\label{eq26}\\ 
Y &=&  \frac{\tilde{\mathcal{U}}}{\bar{R}^2 (1 - 2\tilde{\mathcal{U}})}, \label{eq27}\\ 
Z &=& \rm{ln} (1 - 2\tilde{\mathcal{U}}) -\frac{\tilde{\mathcal{U}}}{1 - 2\tilde{\mathcal{U}}}, \label{eq28}
\end{eqnarray}
Also, at the stellar boundary $r=\bar{R}$, the radial pressure component vanishes, i.e. $p_r(r=\bar{R})=0$ which gives the value of the Bag constant $B_g$ as            
\begin{eqnarray}
B_g &=& \frac{e^{-4 \bar{R}^2 X}}{16 \pi \bar{R}^6}
   \Bigg[2 e^{4 \bar{R}^2 X} \bar{R}^4 + 
    e^{3 \bar{R}^2 X} \bar{R}^4 \Big(-2 + \bar{R}^2 (X - 3 Y)\Big) + 
    128 \alpha_1 Y \Big(-3 + \bar{R}^2 (-9 X + 10 Y) + 
       2 \bar{R}^4 (Y^2  + 7 X Y - 10 X^2) \nonumber\\&& + 
       \bar{R}^6 (42 X^3 - 59 X^2 Y + 12 X Y^2 + Y^3)\Big) - 
    128 \alpha_1 Y e^{\bar{R}^2 X}
       \Big(-6 + \bar{R}^2 (-13 X + 14 Y) + \bar{R}^4 (4 Y^2 + 10 X Y -22 X^2)  \nonumber\\&& + 
       \bar{R}^6 (35 X^3 - 36 X^2 Y + 7 X Y^2 + 2 Y^3)\Big) + 
    128 \alpha_1 Y e^{2 \bar{R}^2 X} \Big\{-3 + 
       \bar{R}^2 (X - Y) \Big(-4 - 2 \bar{R}^2 (2 X + Y) + \bar{R}^4 (X - Y) \times \nonumber\\&&(3 X + Y)\Big)\Big\}\Bigg]\label{eq29}
\end{eqnarray}       
Thus, we have successfully determined the values of $X, Y, Z$ present in the KB metric coefficients, and Bag constant $B_g$ in terms of mass $\mathcal{M}$ and radius $\bar{R}$. From (\ref{eq29}) we see that $B_g$ depends on the parameter $\alpha_1$ while $X, Y, Z$ are independent of $\alpha_1$. Also, in Fig.~\ref{match} we have verified how we smoothly match the metric potentials for both interior and exterior geometries which fulfills the Darmois-Israel condition \cite{Chu:2021uec, darmois1927equations, Israel:1966rt}. By matching these, we obtain the values of the metric constants that characterize our stellar model. 
Now to analyze the physical attributes of our present model we have considered here particularly the compact star 4U 1538-52 with observed mass and radius $M = 0.87 \pm 0.07~M_{\odot},\, \bar{R} = 7.866_{-0.21}^{+0.21}$ km \cite{Rawls:2011jw}. Along with these, we have also particularly taken $\omega = 0.25$ to simplify our calculation. Next, we have calculated the numerical values of the constants $X, Y, Z$ in Table \ref{table1} using the observed values of several candidates for compact stellar objects.\\

\begin{table*}[hbt!]
\centering
\caption{The numerically computed values of the constants $X,\,Y$ and $Z$ for some well known compact stellar objects considering $\omega = 0.25$.}\label{table1}
\begin{tabular}{@{}ccccccccccccc@{}}
\hline
Star & Observed mass & Observed radius & Estimated  & Estimated &  $X$&$Y$&$Z$\\
& ($M_{\odot}$) & (km.) & mass ($M_{\odot}$) & radius (km.)& $\rm{km}^{-2}$ & $\rm{km}^{-2}$ \\
\hline
4U 1538-52 \cite{Rawls:2011jw}& $0.87 \pm 0.07$ & $7.866 \pm 0.21$ & 0.87 & 7.8 & 0.00655890 &0.00403023&-0.644243  \\
SMC X-4 \cite{Rawls:2011jw} & $1.29 \pm 0.05$ & $8.831 \pm 0.09$ & 1.29 & 8.8 & 0.00731424 & 0.00491954 & -0.947383 \\
Vela X-1 \cite{Rawls:2011jw} & $1.77 \pm 0.08$ & $9.56 \pm 0.08$    & 1.77 & 9.5& 0.00883866 & 0.00676124 & -1.407890     \\
Her X-1 \cite{Abubekerov:2008inw}& $0.85 \pm 0.15$ & $8.1 \pm 0.41$ & 0.85 & 8.1 & 0.00564605 & 0.00341692& -0.594622       \\
Cen X-3 \cite{Rawls:2011jw} & $1.49 \pm 0.08$ & $9.178 \pm 0.13$ & 1.49 & 9.2& 0.00767545& 0.00540449& -1.107090 \\
LMC X-4 \cite{Rawls:2011jw} & $1.04 \pm 0.09$ & $8.301 \pm 0.2$ & 1.04 & 8.3 & 0.00669853& 0.0042560 & -0.754658  \\
PSR J1614-2230 \cite{Demorest:2010bx} & $1.97 \pm 0.04$ & $9.69 \pm 0.2$   & 1.97& 9.7& 0.00971519 & 0.00794205 &-1.661370 \\
PSR J1903+327 \cite{Freire:2010tf} & $1.667 \pm 0.021$ & $9.438 \pm 0.03$ & 1.67 & 9.4 &0.00840356&0.00623168&-1.293170           \\
4U 1820-30 \cite{Guver:2008gc} & $1.58 \pm 0.06$&  $9.316 \pm 0.086$  & 1.58 & 9.3&0.00804157&0.00580843&-1.197890 \\
EXO 1785-248 \cite{Ozel:2008kb} & $1.3 \pm 0.2$ & $8.849 \pm 0.4 $ &1.3 &8.85&0.00725186&0.00488178&-0.950337 \\
\hline
\end{tabular}
\end{table*}

\section{Physical characteristics of present model}\label{phy}

In this section, we shall analyze the most important physical parameters of our chosen compact stellar objects through graphical representations as well as by numerical techniques due to highly complicated analytical forms.  The detailed discussions are given in the following subsections:
\subsection{Regularity of our chosen metric}\label{mp}
\begin{figure}[H]
    \centering
        \includegraphics[scale=.55]{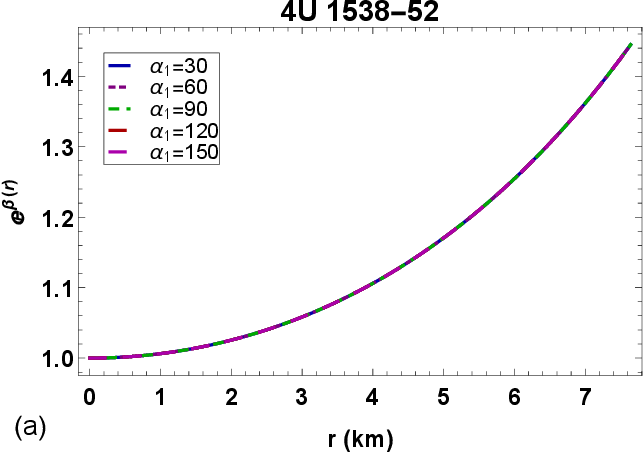}
         \includegraphics[scale=.55]{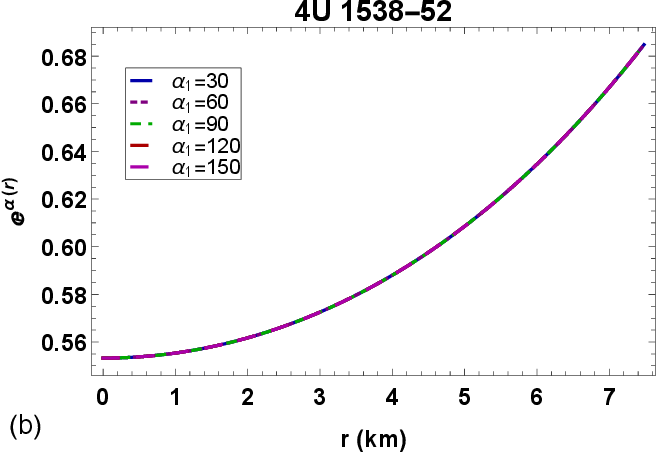}
        \caption{Variation of the gravitational metric functions $e^{\beta(r)}$ and $e^{\alpha(r)}$ with respect to `r'.}\label{metric}
\end{figure}

In this subsection, we discuss the behavior of our chosen metric potential temporal components $e^{\alpha(r)}$ and spatial components $e^{\beta(r)}$.
We can easily verify that $[{e^{\alpha(r)}}]_{r = 0}= e^Z$, a nonzero constant, and $[{e^{\beta(r)}}]_{r=0} = 1$, which confirms that both metric potential components are finite at the center and have regularity throughout the model, $r < \bar{R}$ \cite{Delgaty:1998uy, Pant:2010iub}. Moreover, $\Big[\frac{d(e^{\alpha(r)})}{dr}\Big]_{r=0} = (2Yr{e^{Yr^2 + Z}})\Big\rvert_{r=0} =0$ \text{and}\\ $\Big[\frac{d(e^{\beta(r)})}{dr}\Big]_{r=0} = (2Xre^{Xr^2})\Big\rvert_{r=0} =0$.\\
Thus, we see that at the center of the star, the derivatives of the metric potential components vanish. These components are even positive and consistent within the interior of the star, as seen from the radial profiles of the metric coefficient components shown in Fig.~\ref{metric}. Here we show the regularity of the metric potentials by varying the parameter $\alpha_1$. Thus we verify that the metric potential components are well-behaved within the stellar range $(0, \bar{R})$.

\subsection{Regularity of the fluid components associated with OBM and QM}

\begin{figure}[H]
    \centering
        \includegraphics[scale=.47]{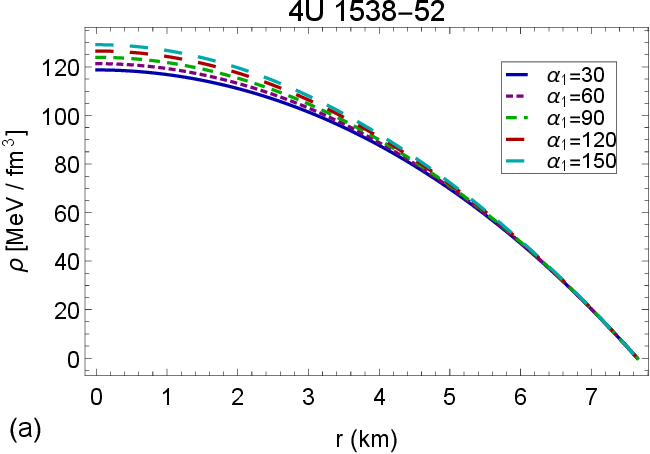}
         \includegraphics[scale=.47]{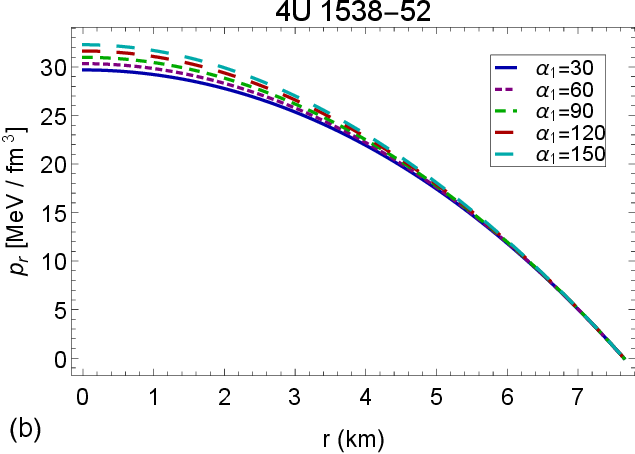}
         \includegraphics[scale=.47]{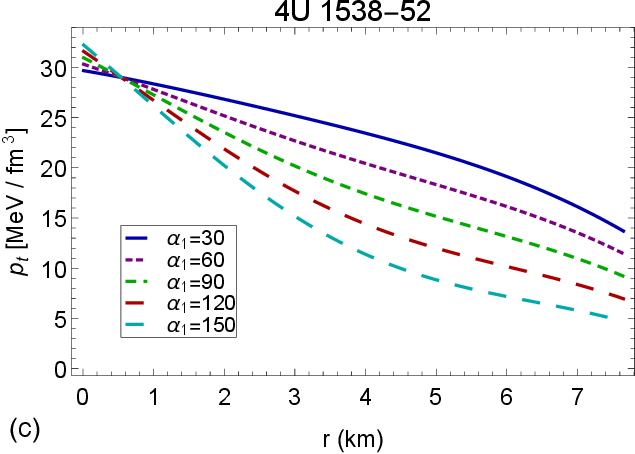}
        \caption{Profiles of baryonic matter-energy density and pressure components with respect to `r'.}\label{rho}
\end{figure}

\begin{figure}[H]
    \centering
        \includegraphics[scale=.47]{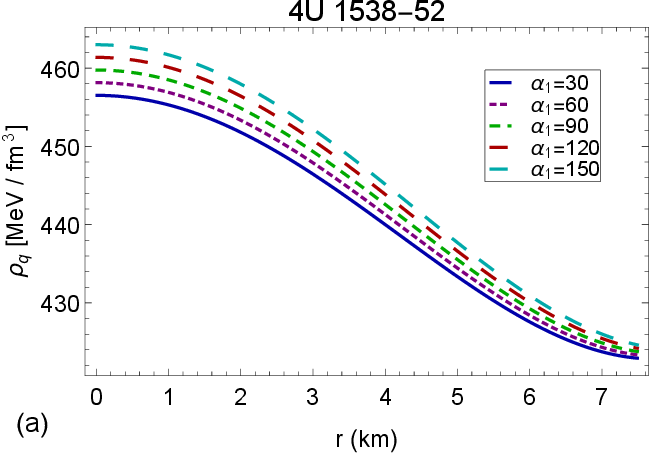}
         \includegraphics[scale=.47]{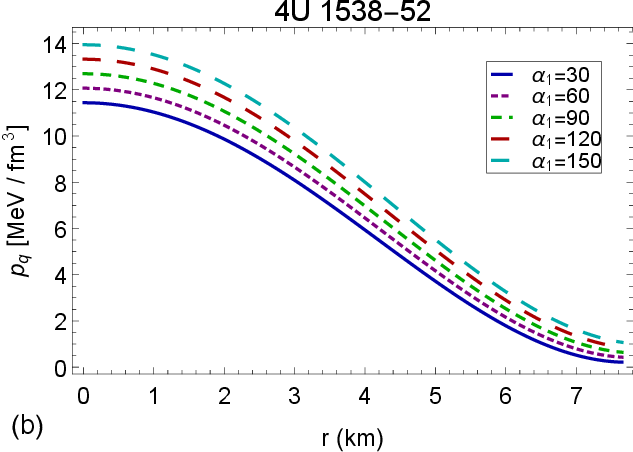}
    \caption{Profiles of QM density $\rho_q$ and QM pressure $p_q$  with respect to `r' with required magnified inset.}\label{rhoq}
\end{figure}

The density of confined matter is very important in establishing how stable a stellar structure is against gravitational collapse, and pressure is important in defining the stellar boundaries and overall stability \cite{chandrasekhar1984stars}. In Fig.~\ref{rho}, we plotted the OBM matter-energy density and pressure components, which indicates that they are all monotonic decreasing functions of radius $r$ with the maximum value at the center of the star. Also, $\rho$ and $p$ are non-negative inside the star. 
We have also displayed the energy density and pressure profiles due to QM in FIG.~\ref{rhoq}. The figure shows that $\rho_q$ and $p_q$ both display positive nature inside the compact stellar object.

We have also provided the numerically computed values of the Bag constant, the effective central fluid components ($\rho_c ^{eff}$, $p_{r_c} ^{eff}$) and the effective surface energy density ($\rho_s ^{eff}$) in Table~\ref{table2} for different values of the parameter $\alpha_1$. From this table, we clearly notice that only the Bag constant decreases with increasing $\alpha_1$ while the other effective parameters increase.

\begin{table}[H]
\centering
\caption{\label{table2} Numerically computed values of the Bag constant, the effective central fluid components as well as the effective surface energy density for the compact object 4U 1538-52 considering $\omega = 0.25$.}
\begin{tabular}{ccccc}
%\begin{tabular}{c|c|c|c|c}
\hline
\hline
          & \multicolumn{4}{c}{4U 1538-52}  \\
\cline{1-5}
$\alpha_1 ~$ & $B_g$ ($\text{km}^{-2}$) & $\rho_c ^{eff} \times 10^{15}$ ($\text{gm}~\text{cm}^{-3}$) & $\rho_s ^{eff}\times 10^{14}$ ($\text{gm}~\text{cm}^{-3}$) & $p_{r_c} ^{eff}\times 10^{13}$ ($\text{dyne}~\text{cm}^{-2}$) \\
\hline
\hline
 30  & 0.000139064    & 1.02274 &  7.51715  & 7.31275 \\
 60  & 0.000138977    & 1.03023 &  7.52390  & 7.53982 \\
 90  & 0.000138889    & 1.03773 &  7.53065  & 7.76689 \\
 120  & 0.000138802    & 1.04523 &  7.53740  & 7.99397 \\
 150  & 0.000138715    & 1.05272 &  7.54415  & 8.22104 \\
\hline
\end{tabular}
\end{table}

\subsection{Nature of the OBM fluid components}
  
\begin{figure}[H]
    %\centering
        \includegraphics[scale=.4]{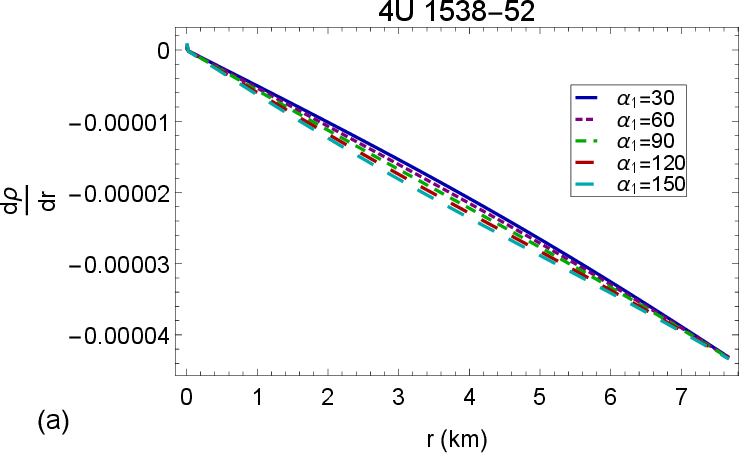}
         \includegraphics[scale=.4]{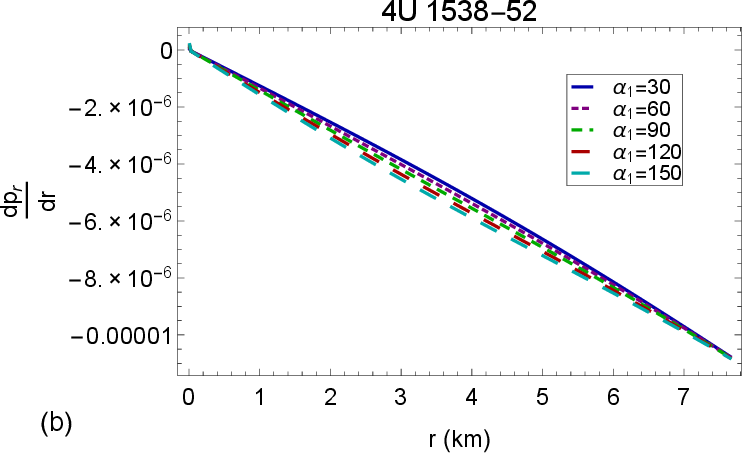}
          \includegraphics[scale=.4]{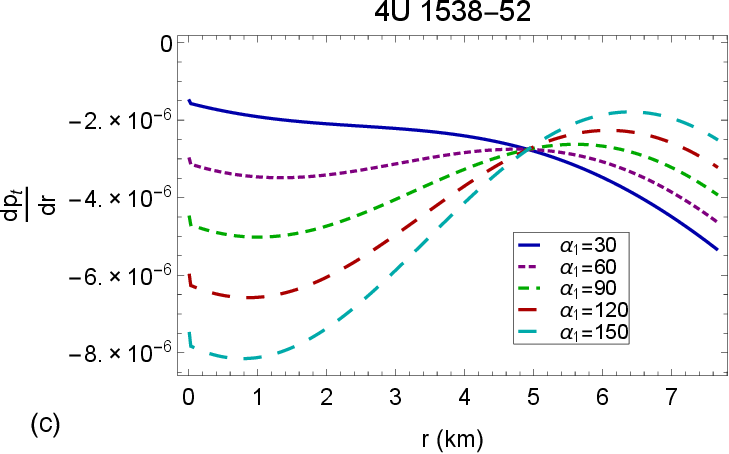}
       \caption{$\frac{d\rho}{dr}$, $\frac{dp_r}{dr}$ and $\frac{dp_t}{dr}$ are plotted against $r$ inside the stellar interior.\label{deri}}
\end{figure}

Here, we shall discuss the variation of density and pressure gradients due to OBM for our present model. Due to the very complicated analytical form, we shall discuss them through graphical presentations.
Thus, from Fig.~\ref{deri}, we can check that the density and pressure gradients due to OBM stay negative throughout the fluid sphere, which is expected for a physically realistic model.

\subsection{Effective mass function}

Since the active stellar mass is gravitationally restricted to a finite spatial extent $(r = R)$, we know that it depends on the energy-density profile and increases with the confining radius \cite{buchdahl1959general, glendenning2012compact}. Misner-Sharp \cite{Misner:1964je} proposed the following formula for the mass of a sphere:
\begin{eqnarray}\label{mma} 
m(r)= \frac{r}{2} \Big(1-g^{\phi \nu }r_{,\phi}r_{,\nu} \Big),
\end{eqnarray}
which leads to
\begin{eqnarray}\label{mmb} 
m(r)= \frac{r}{2} \Big(1- e^{-\beta}\Big).
\end{eqnarray}

Hence, we can easily derive the effective mass function $m^{eff}(r)$ by computing the integral connected directly to the effective energy density (\ref{ef1}) using the following expression:
\begin{eqnarray}
 m^{\text{eff}}(r)=4\pi\int^r_0{\rho^{\text{eff}} r^2 \,dr} =  4\pi\int^r_0{(\rho+\rho_q) r^2 \,dr},~~~\label{mm}
 \end{eqnarray}
After employing the metric potentials on (\ref{mm}) we finally obtain \cite{florides1983complete, kumar2022isotropic}, 
\begin{eqnarray}\label{mm1} 
    m^{\text{eff}}(r)=\frac{r}{2}\Big(1-e^{-\beta(r)}\Big)=\frac{r}{2}\Big(1-e^{-Xr^2}\Big).
\end{eqnarray}

\begin{figure}[H]
    \centering
        \includegraphics[scale=.465]{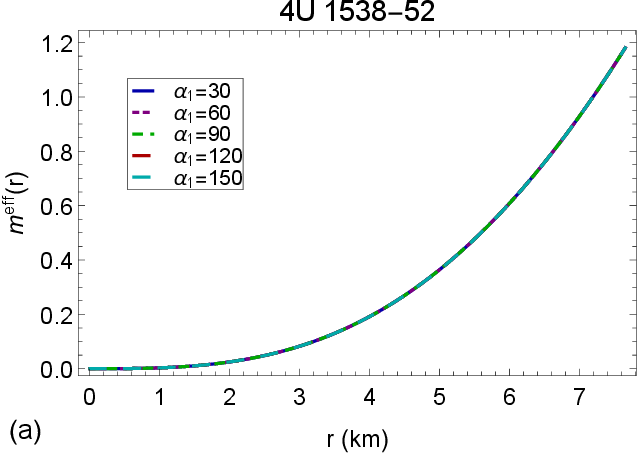}
        \includegraphics[scale=.465]{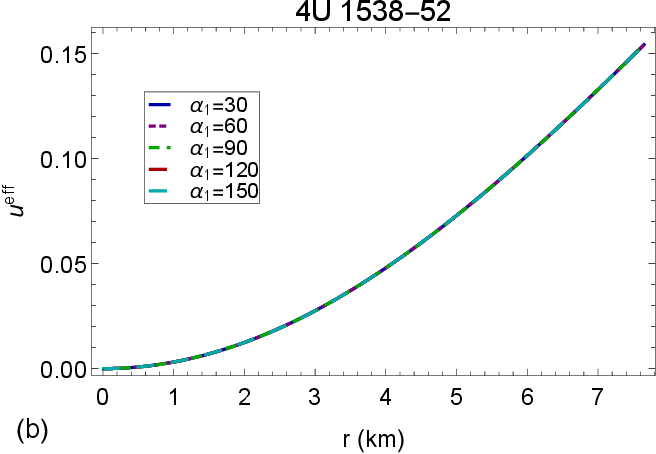}
        \includegraphics[scale=.465]{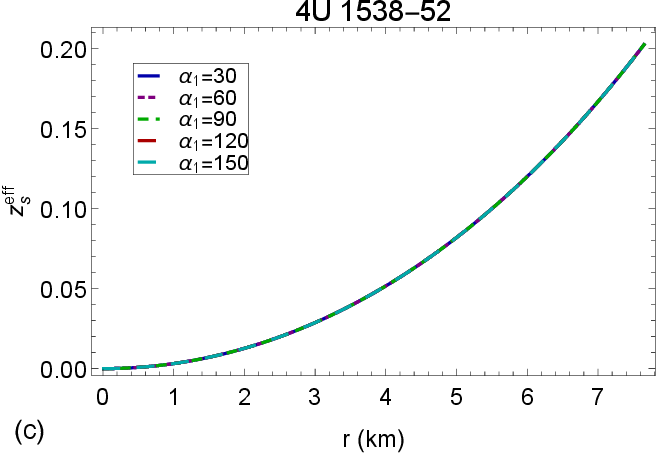}
       \caption{(a) Effective mass function, (b) effective compactness factor, and (c) effective surface redshift are plotted against `r'.}\label{mass}
\end{figure}

It should be noted that the effective mass function $m^{\text{eff}}(r)$ is a function of radius $r$. Furthermore, it is evident that $m^{\text{eff}}\to  0$ as $r \to 0$, which means the effective mass function is finite at the center of the fluid sphere. The variation of mass function (\ref{mm}) has been plotted against $r$ in Fig.~\ref{mass}.\par
Clearly, effective mass is regular at the center as it is directly proportional to the radial distance $r$ and maximum mass is attained at the surface $r=\bar{R}$ as displayed in Fig.~\ref{mass}.\par
\subsection{Effective compactness factor}

Furthermore, the effective compactness factor of a celestial object is determined by a dimensionless parameter $u^{\text{eff}}(r) = \frac{m^{\text{eff}}(r)}{r}$. The graphical evolution of the effective compactness factor $u^{\text{eff}}(r)$ has been analyzed in Fig.~\ref{mass} and shows that $u^{\text{eff}}(r)$ is monotonically increasing with $r$. 

\subsection{Effective surface redshift}

Now the effective surface redshift $z_s ^{\text{eff}}(r)$ for the present compact star candidate can be obtained by using the expression of effective compactness factor $u^{\text{eff}}(r)$ given by $z_s ^{\text{eff}}(r)=\frac{1}{\sqrt{1-2u^{\text{eff}}(r)}}-1$. We have shown the graphical evolution of $z_s ^{\text{eff}}(r)$ in Fig.~(\ref{mass}) from center to surface. Clearly, the effective surface redshift depends on the stellar mass and radius, in other words, on the surface gravity.

\section{Stability analysis of our model}\label{stable}

\subsection{Causality condition via Herrera's cracking method}

\begin{figure}[H]
    \centering
        \includegraphics[scale=.47]{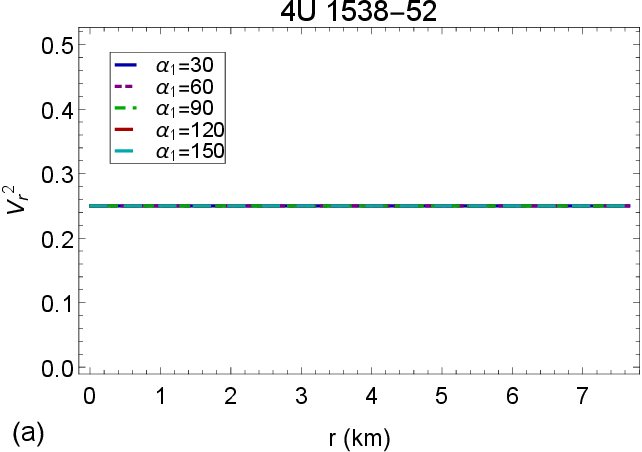}
        \includegraphics[scale=.47]{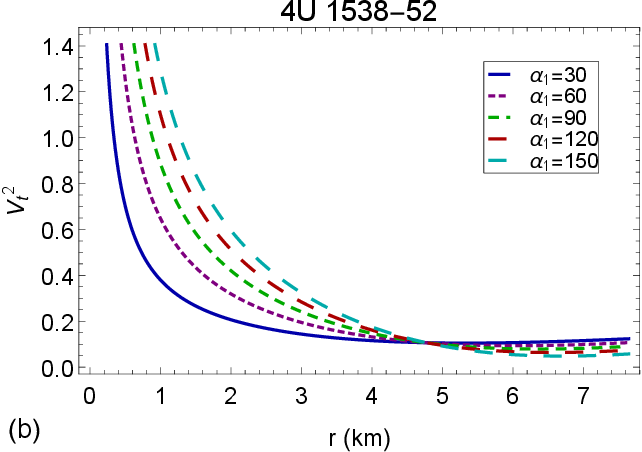}
        \includegraphics[scale=.47]{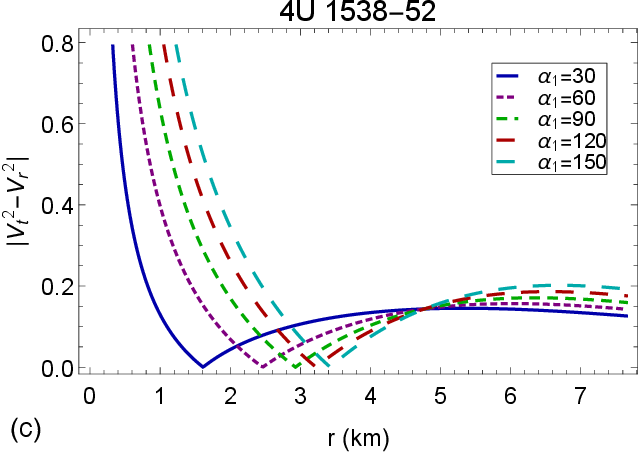}
        \caption{Visualization of sound velocity components and stability factor against `r'.}\label{sv}
\end{figure}

This section will now cover the causality criterion, which is a crucial "physical acceptability condition" for realistic models. Herrera's cracking technique and sound velocity components will be used in this discussion. We first discuss the causality requirement of our model, which states that for a physically realistic model, the square of sound velocity $V^2$=$\frac{dp}{d\rho}$ should be less than unity \cite{Herrera:1992lwz, Abreu:2007ew}. This means that the speed of sound does not exceed the speed of light. Thus, using the expressions~(\ref{f1})-(\ref{f3}), we derive the radial and tangential sound speed components for our anisotropic model as follows:
\begin{eqnarray}
V_r^2&=&\frac{dp_r}{d\rho} ~~\rm{and} \label{sp1}\\
V_t^2&=&\frac{dp_t}{d\rho}.\label{sp2}
\end{eqnarray}
Due to the complexities in their analytic expressions, we evaluated them graphically in Fig.~(\ref{sv}) and found that both are confined within the intended range $[0, 1]$ inside the stellar object. This is often referred to as the causality condition.

It is evident from the figure that the sound velocity components are always positive, regardless of the density of matter. Therefore, our proposed hybrid model satisfies the causality condition.\\
In addition, Herrera developed the "cracking" (or overturning) technique \cite{Herrera:1992lwz} for relativistic compact objects subjected to minor radial perturbations. Furthermore, Abreu et al. used the cracking concept in their investigation \cite{Abreu:2007ew} and proposed the concept of stability factor. It is mathematically described as $|V_t^2-V_r^2|<1$ and its profile is given in Fig.~(\ref{sv}). Here we can clearly assess from the figure that this criterion is met throughout our model. Hence, our model is physically well consistent and potentially stable throughout the stellar distribution because it obeys the causality condition as well as Herrera's cracking concept.

\subsection{Energy conditions}

There are certain mathematical constraints that must be fulfilled by the stress-energy tensor to deal with a physically realistic and feasible matter field. These constraints are generally known as energy conditions (EC). These ECs are invariant in terms of coordinates. These conditions play a vital role in assessing the normal and unusual nature of matter within a stellar structure model. As a result, these conditions have attracted a lot of focus in the study of cosmological phenomena. These conditions are usually termed as (i) Null energy condition (NEC), (ii) Weak energy condition (WEC), (iii) Strong energy condition (SEC), and (iv) dominant energy condition (DEC) \cite{bondi1947spherically, witten1981new, visser1997energy, Andreasson:2008xw, garcia2011energy}. These conditions are fulfilled if the inequalities listed below hold true at every point of the fluid sphere:
\begin{itemize}
\item \textbf{NEC}:~$\rho+ p_r  \geq 0,~\rho+ p_t  \geq 0$, 
\item \textbf{WEC}:~$\rho+ p_r  \geq 0,~\rho + p_t  \geq 0,~ \rho   \geq 0$, 
\item \textbf{SEC}:~$\rho+ p_t \geq 0,~\rho + p_r + 2p_t \geq 0$, 
\item \textbf{DEC}:~$\rho-|p_r|  \geq 0,~ \rho-|p_t|  \geq 0,~\rho \geq 0$.
\end{itemize}

\begin{figure}[H]
    \centering
        \includegraphics[scale=.465]{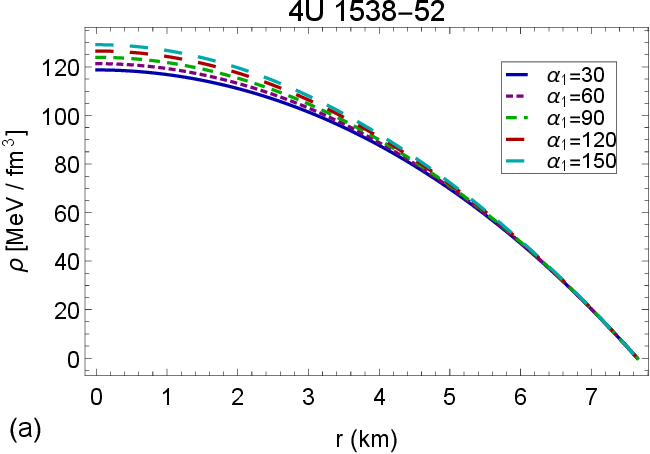}
        \includegraphics[scale=.465]{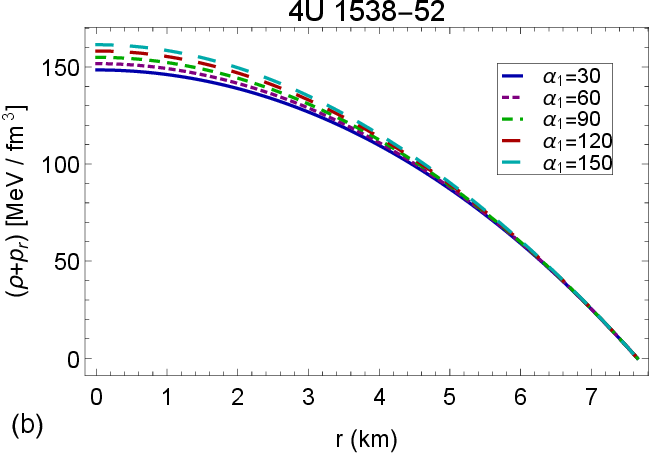}
        \includegraphics[scale=.465]{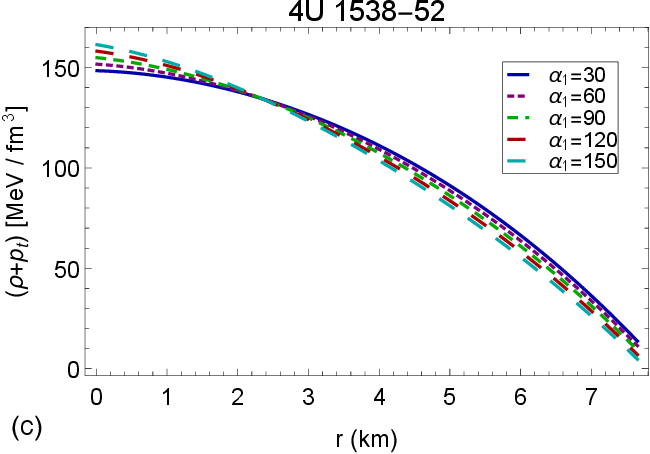}
        \includegraphics[scale=.465]{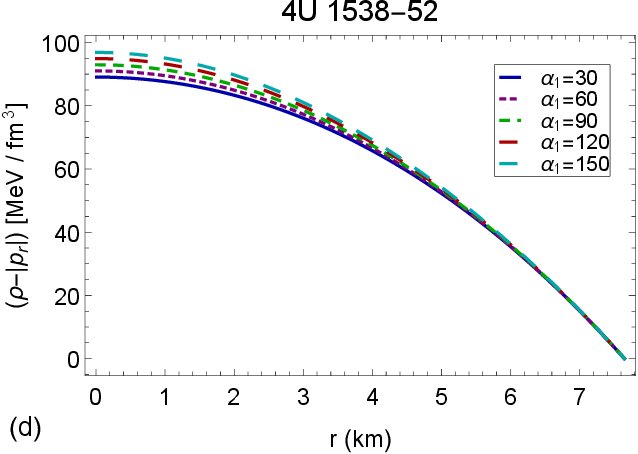}
        \includegraphics[scale=.465]{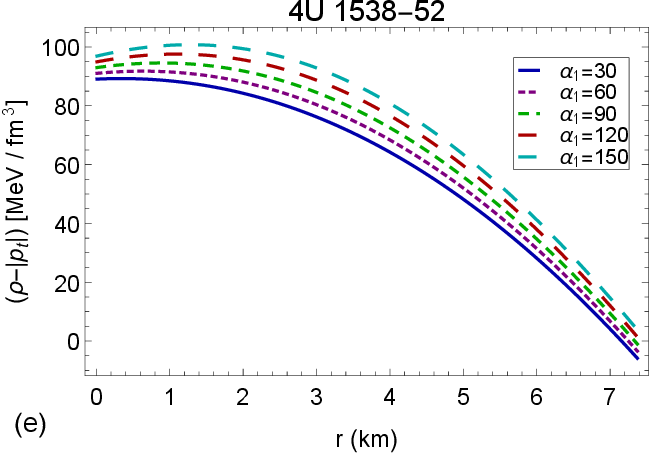}
        \includegraphics[scale=.465]{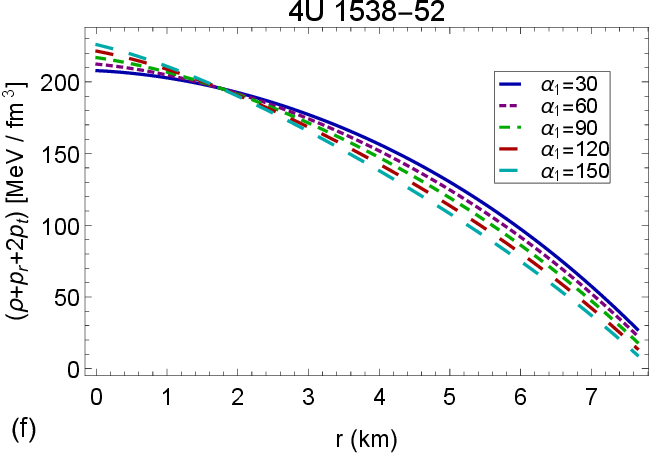}
        \caption{Evolution of energy conditions for the present model.}
    \label{ec}
\end{figure}

Now we shall investigate whether these inequalities hold. So for this, we have plotted the above bounds of all energy conditions in Fig.~(\ref{ec}) and we found that our present hybrid model satisfies all these conditions completely at every point inside the fluid sphere. Hence, our model is physically acceptable.
 
\section{Concluding Remarks}\label{con}
On the basis of cosmological observations, it has been determined that our universe has two phases of accelerated expansion: cosmic inflation in the early universe and acceleration in the current expansion of the Universe. Scientists have been looking at the current cosmic expansion and the nature of DE. Numerous efforts have been made to modify the GR based on various strategies. One of such modifications to GR is $f(\mathcal{G})$ gravity.

Now, the problem of determining a suitable model for the realistic geometry of interior compact objects has drawn attention in both GR and extended theories of gravity such as $f(\mathcal{G})$ gravity. In the current article, we developed a method to investigate the possible formulation of a hybrid stellar model in the context of modified $f(\mathcal{G})$ theory, which is one of the extensions of GR. It is challenging work to model such an astrophysical object using this theory. For this purpose, we investigate the specific compact object 4U 1538-52 by considering a well-known power law model: $f(\mathcal{G}) = \alpha_1 \mathcal{G}^{n_1}$. In this literature, to the best of our knowledge, for the first time we have investigated a hybrid stellar model with an anisotropic fluid distribution under $f(\mathcal{G})$ gravity due to its complicated structure combined with both OBM and QM. By comparing the interior metric with the widely recognized exterior metric, the analytical solution for $f(\mathcal{G})$ gravity has been found. According to the physical analysis of the results, this anisotropic hybrid stellar model in $f(\mathcal{G})$ gravity has the following conclusive properties:

\begin{itemize}
    \item The Darmois-Israel condition has been fulfilled by smoothly matching the metric potentials for both the interior and exterior geometries.
    \item The chosen KB metric is regular throughout the stellar interior.
    \item At the center, baryonic matter-energy density and pressure components reach their maximum values. These are decreasing functions as an outcome.
    \item The energy density and pressure profiles due to QM both exhibit positive features within the compact stellar object.
    \item In Table~\ref{table2}, the values of the Bag constant and other effective parameters have been derived using the numerical technique. It is evident that when $\alpha_1$ increases, only the Bag constant decreases, whereas the other effective parameters experience an increase. 
    \item The pressure and density gradients generated by OBM remain negative throughout the fluid sphere, which one would anticipate from a physically accurate model.
    \item The effective mass function is regular in the center, as it is directly proportional to the radial distance $r$, and the maximum mass is obtained on the surface.
    \item The effective compactness factor and the effective surface redshift increase monotonically with the radial distance $r$.
    \item The causality condition is met since the radial and transverse sound speeds remain within the bound $[0, 1]$ inside the stellar object. This model also satisfies Herrera's cracking criterion. As a result, our model is physically consistent and potentially stable across the stellar distribution in $f(\mathcal{G})$ gravity.
    \item All energy conditions are met, indicating a realistic matter content in $f(\mathcal{G})$ gravity.
\end{itemize}
As a result of all the significant findings, we arrive at the conclusion that we can build a physically acceptable, stable, and singularity-free generalized hybrid stellar model throughout the interior fluid distribution in this specific $f(\mathcal{G})$ gravity model. Therefore, we can state that the various physical properties of strange star objects can be examined at both theoretical and astrophysical gauges by means of an extremely dense, compact stellar object composed of QM. Also in the study of compact stars, $f(\mathcal{G})$ gravity has been very appealing in recent years. To the best of our knowledge, the existence and study of various astrophysical objects and particle physics within their highly dense cores motivated researchers to seek more authentic solutions to field equations. Even more exciting would be if theoretical and observational research could lead us in the right direction toward modifying GR to make it consistent with the standard model of particle physics. Therefore, we expect that this hybrid stellar model will contribute to the larger-scale astrophysical scenario.

\section*{CRediT authorship contribution statement}
\textbf{Pramit Rej}: Conceptualization, supervision, Validation, Methodology, Software, Writing - original draft, Investigation, Writing - review \& editing.

\section*{Acknowledgements} Pramit Rej is thankful to the Inter-University Centre for Astronomy and Astrophysics (IUCAA), Pune, Government of India, for providing Visiting Associateship.

\section*{Declarations}
\textbf{Funding:} The author did not receive funding in the form of financial aid or grant from any institution or organization for the present research work.\par
\textbf{Data Availability Statement:} The results are obtained
using purely theoretical calculations and can be verified analytically;
therefore, this manuscript does not have associated data, or the data will not be deposited. \par
\textbf{Conflicts of Interest:} The author has no financial interest or involvement that is relevant by any means to the content of this study.

\bibliographystyle{apsrev4-1}
\bibliography{hybrid.bib}

\end{document}